\documentclass[5p]{elsarticle}
\usepackage{hyperref}
\usepackage{amsmath}
\usepackage{url}
\begin{document}
\begin{frontmatter}
\title{Influence of the multiple scattering of relativistic electrons on the line width of the backward Parametric X-ray Radiation in the absence of photo absorption}
\author{Tabrizi Mehdi}
\address{Parallel Computing Laboratory (2216), Department of Physics, Faculty of Science, Razi University of Kermanshah, 67149-67346, Iran}
\ead[url]{http://fm.razi.ac.ir/tabrizi}
\ead{tabrizi@bk.ru, tabrizi@razi.ac.ir}
\cortext[cor1]{Fax: +98 (0833) 4274556}
\begin{abstract}The multiple scattering effect on the line width of the backward Parametric X-ray Radiation (PXR) in the extremely Bragg geometry, produced by low energy relativistic electrons traversing a single crystal, is discussed. It is shown that there exist conditions, when the influence of photo absorption on the line width can be neglected, and the only multiple scattering process of relativistic electrons in crystal leads to the PXR  lines' broadening. Based on the obtained theoretical and numerical results for the line width broadening, caused by the multiple scattering of  $30$ and $50$ MeV  relativistic electrons in a Si crystal of varying thicknesses, an experiment could be performed to help to reveal the scattering effect on the PXR lines in the absence of photo absorption. This leads to a more accurate understanding of the influence of scattering phenomenon on the line width of the backward PXR and helps to a better construction of a table-top narrow bandwidth X-ray source for scientific and industrial applications.
\end{abstract}
\end{frontmatter}
\section{Introduction}
Parametric X-ray Radiation (PXR) is produced when relativistic electrons fall at a small incidence angle with respect to one of the crystallographic atomic planes (see \cite{TM,ARTRU,BAR} and references therein). Recently, production of PXR by ultra-relativistic protons was observed in a bent crystals \cite{SCAND}. This radiation is mainly concentrated in directions close to the Bragg's angles of particle's field reflection from such planes. The case of PXR in the "extremely Bragg geometry" or "backward" PXR is of special interest, when relativistic electrons fall onto a crystal at a small angle $\psi>\psi_{c}$ with respect to one of the crystallographic axes ($z$-axis in Fig. \ref{fig1}), since in this case, the contribution of bremsstrahlung and channeling radiation are then considerably suppressed. Here $\psi_{c}=\sqrt{4Ze^{2}/Ed}$ is the critical angle of axial channeling \cite{LIN}, $d$ - the inter atomic distance along the $z$-axis, $E$ - particle energy. \\
The backward PXR in extremely Bragg geometry was observed at low \cite{FREUD1} and high \cite{BACKE1} energies of relativistic electrons, but the line width was measured in \cite{BACKE1}. Measurements of the line width of the backward PXR at high energies of electrons \cite{BACKE1} show that the PXR has a very narrow line width of a few meV (milli-electron-Volt) which points to the fact that such a quasi-monochromatic and narrow bandwidth X-ray radiation can be used in many applications \cite{SONES, YUM,TAKA}. Such narrow lines appear in the spectral angular radiation density as a result of the interference of reflected waves from crystallographic atomic planes, oriented perpendicular to the $z$-axis (Fig. \ref{fig2}). 
The natural width of these lines is determined by the number of crystallographic planes which the relativistic electrons interact with. However, the results of experiment \cite{BACKE1} show that the line width of the backward PXR is much larger than its natural line.\\
There are two kinds of effects which destructively influence on PXR lines. The "instrumental" effects such as finite detector opening \cite{FREUD}, finiteness of the collimator \cite{GOGOL} and angular spread of the electron beam \cite{BACKE} are connected with the instruments for measurement of PXR lines.\\
The "non-instrumental" effects on PXR lines, caused by such processes as absorption of radiated photons and the multiple scattering of relativistic electrons in crystal are connected with the physical phenomena while producing PXR. In the subsequent discussion, the two aforementioned "non-instrumental" effects on the line width of the backward PXR will be considered .\\
For thick crystals, such that their thickness $L$ is larger than the absorption length $L_{a}$ of radiated photons, the absorption effect plays the leading role in the formation of PXR lines \cite{BACKE1,NITTA}. Although, at $L\gg L_{a}$, PXR intensity ultimately no longer increases with L \cite{BAR1}.
The multiple scattering process of relativistic electrons in crystal is the another physical phenomena which makes destructive contribution to the line width of PXR \cite{BACKE1,FREUD,BACKE} (see Fig. \ref{fig1}). However, this process makes itself more evident in the formation of PXR lines, if low energy relativistic electrons traverse a crystal of the thickness $L$ less than $L_{a}$.\\
In this paper, the influence of the small angle multiple scattering of relativistic electrons of different energies on the line width of the backward PXR in extremely Bragg geometry in a single crystal of varying thicknesses $L$ ($L<L_{a}$) under the conditions of  absence of photo absorption, is investigated. The conditions, when the absorption effect of radiated photons is negligible, are found. 
In what follows, we shall use the system of units, in which $c=\hbar=1$.
\section{Spectral angular density of the backward PXR including the small angle multiple scattering process \label{sec2}}
Analysis of the experimental data on the backward PXR shows \cite{BACKE1} that, apart from the absorption, the important contribution to the characteristics of the PXR lines, such as line width, is made by deflection of relativistic electrons from initial straight forward direction in crystal. In other words, the assumption of the particle's straight trajectory and its \textit{no}-effect on the line shape is inconsistent with the experimental data.\\
\begin{figure}
  \centering
  \includegraphics[width=\linewidth]{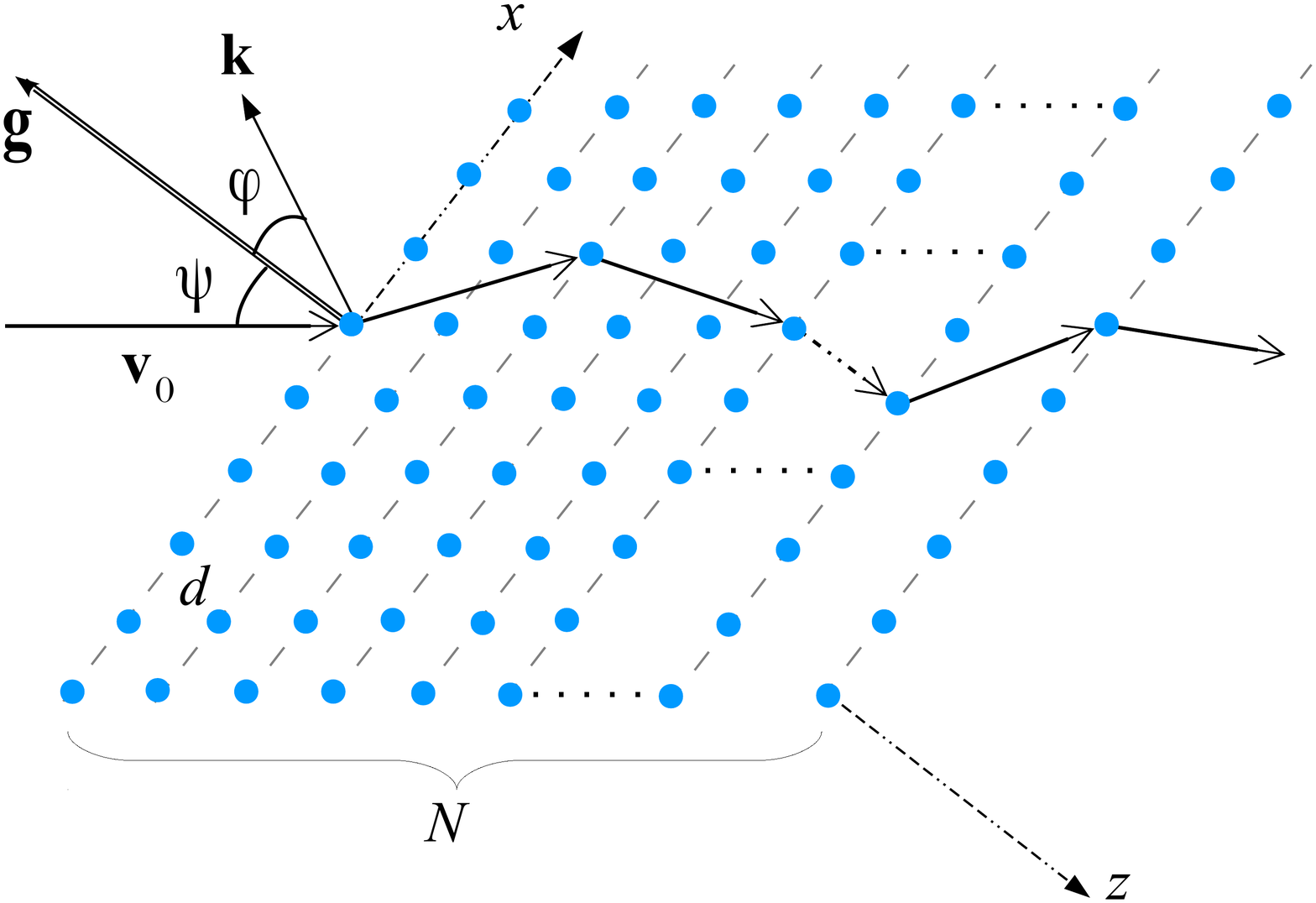}
  \caption{2D Schematic representation of the multiple scattering of relativistic electrons in a crystal producing the backward PXR in extremely Bragg geometry (view from the $y$-axis): $\mathbf{v}_{0}$ - velocity vector of the incoming electron, $\mathbf{g}$ - reciprocal lattice vector, $\mathbf{k}$ - wave vector of the radiated photons, $\psi$ - incidence angle of electrons with respect to the $z$-axis, $\phi$ - angle between $\mathbf{k}$ and $\mathbf{g}$,  $N$ - number of the crystal planes, $d$ - distance between lattice planes.}
  \label{fig1}
\end{figure}
For a relativistic electron ($E\gg m_{e}$), the variation of its velocity $\mathbf{|\dot{v}|}$ due to scattering is small, since  $\mathbf{|\dot{v}|}\sim1/E$ \cite{SHULGA1}. Therefore, for relativistic electrons, falling onto a crystal at a small angle $\psi>\psi_{c}$, one can represent the velocity $\mathbf{v}(t)$ in the form \cite{SHULGA1}
\begin{equation}
\mathbf{v}(t)\approx\mathbf{v}_{0}(1-\frac{1}{\rm{2 v_{0}^{2}}} \rm{v_{\perp}^{2}})+\mathbf{v}_{\perp}(\textit{t}),
\label{eq0}
\end{equation}
where $\mathbf{v}_{0}$ - velocity of the incident electrons, $\mathbf{v}_{\perp}(t)$ - component of $\mathbf{v}(t)$ which is perpendicular to $\mathbf{v}_{0}$ and $\rm v_{\perp} \ll v_{0}$. The latter inequality allows us to use the so-called small angle scattering approximation $\theta\sim \rm (v_{\perp}/v)$. As it is shown in Fig. \ref{fig2}, $\mathbf{v}_{\perp}(t)$ has two components $\mathbf{v}_{\perp}(t)=\rm (v_{x},v_{y})$.\\
The deviation of trajectory of relativistic electrons from the straight forward direction in a crystal, is mainly connected with the multiple scattering process \cite{SHULGA} and the channeling phenomenon \cite{LIN}. In the following, it will be assumed that the relativistic electron impinges to one of the crystallographic axes of a crystal ($z$-axis in Fig. \ref{fig1}) at an angle $\psi$, which is larger than the critical angle of axial channeling $\psi_{c}$. Therefore, the channeling of relativistic electrons in the crystal does not take here place.\\
\begin{figure}
  \centering
  \includegraphics[scale=0.32]{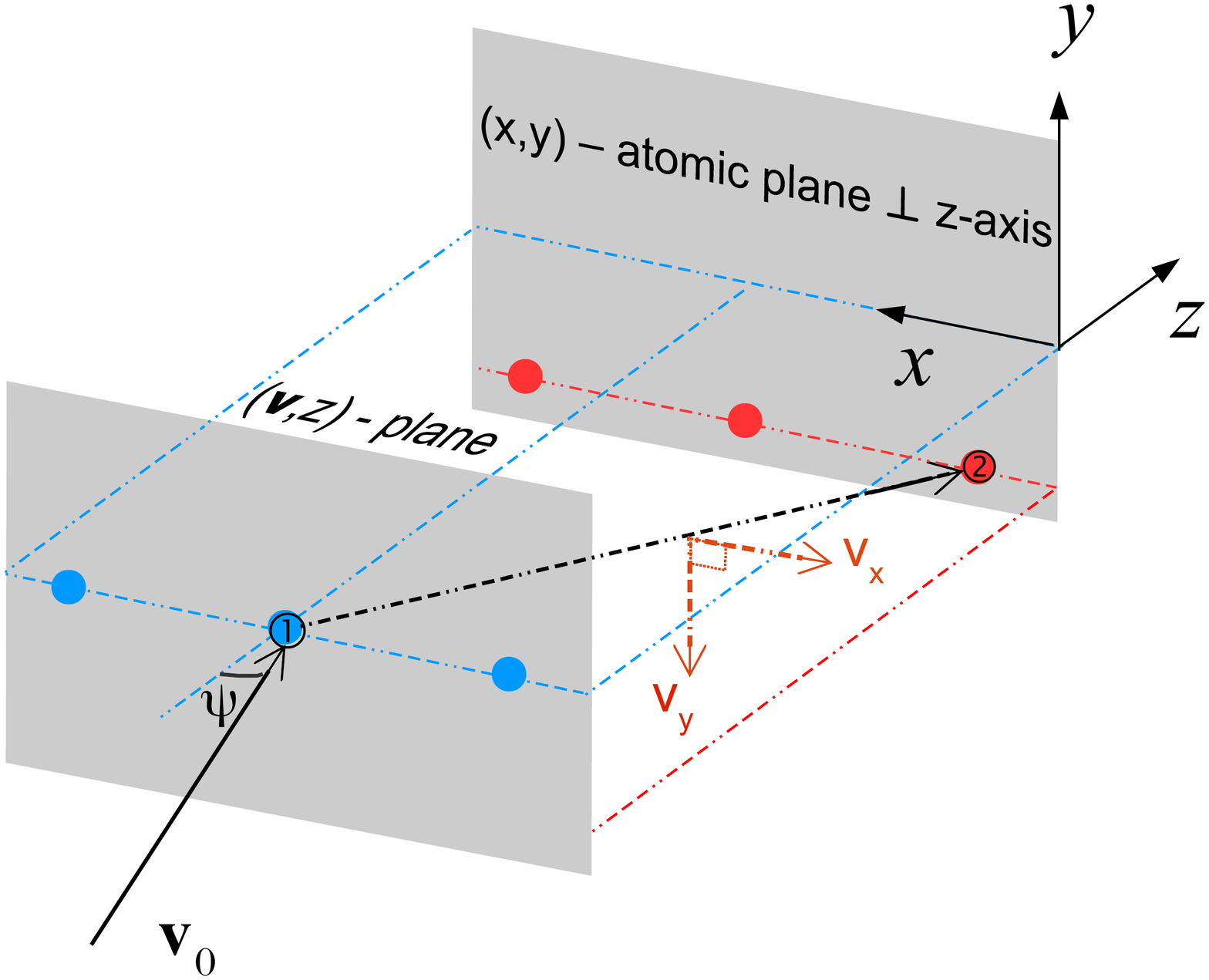}
  \caption{3D schematic representation of a single scattering of electron on the atom 1 of the top plane going to the atom 2 of the middle plane along the $x$- and $y$-axes.}
  \label{fig2}
\end{figure}
As it is shown in Fig. \ref{fig1}, the small angle scattering of relativistic electrons during the passage through the crystal leads to the small change of mean free path of particles between sequential collisions with crystallographic atomic planes. This, in its turn, leads to the destruction of interference of reflected waves from the various planes. One of the important features of scattering process of relativistic electrons at $\psi>\psi_{c}$, is that the process is asymmetric in $(x,y)$-plane (Fig. \ref{fig2}), namely, the scattering along the $x$-axis in $(\mathbf{v},z)$-plane and along the $y$-axis in $(x,y)$-plane essentially are different. The reason is that the correlations between subsequent collisions of relativistic electrons with atomic chains, parallel to the $z$-axis are of substantial. Due to these correlations, the asymmetric scattering occurs.\\
The correlated scattering takes place mainly along the azimuthal angle in $(x,y)$-plane, perpendicular to the $z$-axis\cite{SHULGA1}. A redistribution of relativistic electrons occurs over this angle due to the multiple scattering by different atomic chains. At $\psi>\psi_{c}$, the scattering process can be approximated by a Gaussian one \cite{SHULGA}. The mean square of the multiple scattering angle is then given by $\overline{\theta_{c}^{2}}=q_{c} L$, where $q_{c}$ is the mean square of the correlated scattering angle per unit length. The quantity $q_{c}$ differs from the corresponding one in an amorphous medium by a factor of the order of $R/4\psi a$ \cite{SHULGA}, where $R$ is the Thomas-Fermi radius of screening of the potential of a separate crystal atom.\\
The non-correlated scattering of relativistic electrons occurs on thermal vibrations of atoms in the crystal, and the mean square scattering angle in this case is given by $\overline{\theta_{a}^{2}}=q_{a} L$ of that of an amorphous medium \cite{SHULGA1}. \\
Thus, the deviation of the particle's trajectory along the $y$-axis, which is oriented perpendicular to the $(\mathbf{v},z)$-plane, is caused by both correlated and non-correlated scattering in the crystal, and the mean square scattering angle is given by $\overline{\theta_{y}^{2}}=q_{y} L$, where $q_{y}=q_{c}+q_{a}/2$. The deviation of particle's trajectory along the $x$-axis is mainly caused by non-correlated scattering with mean square scattering angle of $\overline{\theta_{x}^{2}}=q L$, where $q=q_{x}/2$. The $q_{x}$ value differs by less than $10\div15\%$ from the corresponding value for an amorphous medium $q_{x}^{a}$ \cite{SHULGA1}. In the multiple scattering theory of high energy electrons in an amorphous medium, the value $q_{x}^{a}$ is defined by the relation $q_{x}^{a}=\eta (13.6 \,\mathrm{MeV}/E)^{2}/L_{R}$ \cite{BERINGER}, where $L_{R}$ is the radiation length of relativistic electrons and $\eta=(1+0.038\ln\frac{L}{L_{R}})^{2}$. In section \ref{sec4}, we will use $q_{x}^{a}$ instead of $q_{x}$ for numerical calculations.\\
In \cite{TABRIZI}, the spectral angular density of the backward PXR in extremely Bragg geometry, averaged on the multiple scattering process by the method of functional integration \cite{YAGLOM} was obtained. Thus, the appropriate formula with some modifications for the spectral angular radiation density near the lines with the energy $\omega_{n}\approx g/2$ reads as the following
\begin{equation}
\frac{d^{2}W}{d\omega d\Omega}=C\Phi(\phi,\psi)\,L^{2} F(L,\Delta\omega),
\label{eq1}
\end{equation}
where $C=\frac{e^{2}\omega_{n}^{2}|\varepsilon_{\omega{n}}^{\prime}|^{2}}{4\pi^{2}d^{2}}$, $\Phi(\phi,\psi)=\frac{(\phi-\psi)^{2}}{[\gamma^{-2}+(\phi-\psi)^{2}]^{2}}$, $e$ - charge of particle, $\varepsilon^{\prime}_{\omega_{n}}$ - Fourier component of material part of the medium permittivity $\varepsilon_{\omega_{n}}=1+\varepsilon^{\prime}_{\omega_{n}}$, $g=2\pi n/a$, $n$ - integers, $\gamma$ - Lorentz factor of particle, and $F(L,\Delta\omega)$ is defined as 
\begin{equation}
\begin{split}
F(L,\Delta\omega)=2\int_{0}^{1} dz \int_{0}^{z} dy \cos[2 y L \Delta\omega]\\ \times \exp(-\alpha^{2}\sigma_{x}^{2}y^{2}z), 
\end{split}
\label{eq2}
\end{equation}
where $\Delta\omega=\omega-\omega_{n}$, $\alpha=\phi_{r}/\sqrt{2 q_{x}L}$, $\phi_{r}=\phi-\psi$ and $\sigma_{x}=\omega_{n}q_{x}L^{2}$. The parameters $\alpha$, $\sigma_{x}$ and function  $F(L,\Delta\omega)$  are dimensionless. Other  relevant quantities are defined in Fig. \ref{fig1}.\\
In Eq. (\ref{eq1}), $C$ and $\Phi(\phi,\psi)$ do not depend on the multiple scattering parameter $q_{x}$, but $F(L,\Delta\omega)$ does and describes the behaviour of line width of the backward PXR.
Formula (\ref{eq2}) has been obtained in the approximation in which the main contribution to the line width originates from the non-correlated multiple scattering of relativistic electrons in the crystal, i.e. when $\alpha\gg 1$ and $\sigma_{x}\gg 1$.\\
The factor $\alpha\sigma_{x}$  in $F(L,\Delta\omega)$ stands for influence of the multiple scattering of electrons on PXR lines. If $\alpha\sigma_{x}\rightarrow 0$, which means $q_{x}\rightarrow 0$, then the influence of non-correlated multiple scattering process on line width of the backward PXR can be ignored and $\Delta\omega$ in this case is given by $\Delta\omega\sim 1/L$ to within the order of magnitude. At $\sigma_{x}\gg1$ and $\alpha\gg1$, the scattering process leads to the line broadening of the backward PXR and $\Delta\omega$ is defined to within the order of magnitude as
\begin{equation}
\Delta\omega\sim\alpha\sigma_{x}/L ,
\label{eq3}
\end{equation}
which is much larger than the natural line width for the same crystal thickness. 
\section{Necessary conditions for neglecting the photo absorption effect on the line width}
In experiment \cite{BACKE1}, the backward PXR with the photon energies of several KeV was produced by relativistic electrons of the energy of $E\approx855\,$ MeV in a Si crystal of the thickness of $525\mu m$. For such a thick Si crystal ($L>L_{a}$) besides the scattering process, the another important one, causing the line broadening of the backward PXR, is the photo absorption. Both processes exert a parasitic influence on the line width and were experimentally investigated in \cite{BACKE1}. The interpretation of the experimental data for  the multiple scattering effect of electrons in the crystal on the line width of the backward PXR was explained in \cite{BACKE1} on the basis of \cite{BACKE2}. In the work \cite{BACKE2}, the PXR amplitude was first obtained without any directional changes of electrons in the crystal, and then, was modified for small angle multiple scattering. At the end, the corresponding amplitude was calculated using the probabilistic methods.\\
But there is the another point of view. On one hand, the multiple scattering of electrons during the passage through the crystal is a usual parasitic process and occurs along the whole path of charged particles traversing the crystal.  From the other hand, PXR is produced by reflection of relativistic electron's field from all of the crystal planes, even from the last layers. Consequently, the scattering process of electrons destructively influences on the line width of the backward PXR from the first planes of the crystal traversed by electrons till the last points, when the particles leave it. Therefore to obtain  (\ref{eq2}),  the directional changes of electrons in crystal were from the outset included in the PXR spectral angular density, and afterwards, the latter was averaged on the random scattering process.\\
However, Eq. (\ref{eq2}) has been obtained without taking into account the absorption effect on the line width of the backward PXR. In order to find out the conditions, under which it is possible to neglect the influence of the photo absorption on the line width and to put the formula (\ref{eq2}) into practice, we consider the following.\\
From the one hand, the factor $\alpha\sigma_{x}$ in (\ref{eq2}) is proportional to $L$ as $\alpha\sigma_{x}\propto L\sqrt{L}$. Therefore,
according to the formula (\ref{eq3}), the line width of the backward PXR decreases as $\sqrt{L}$ when decreasing the thickness. 
From the other hand, $\alpha\sigma_{x}$ is proportional to the energy of charged particles as $\alpha\sigma_{x}\propto 1/E$. Hence, if we choose thin crystals and decrease the energy of charged particles, but retaining the latter in relativistic domain, then it is possible to fit such values for $L$ and $E$, so that the following conditions can be fulfilled:
\begin{equation}
L<L_{a}\,, \qquad \alpha\sigma_{x}\gg 1.
\label{eq4}
\end{equation}
In other words, at the fulfilment of the conditions (\ref{eq4}), the influence of the photo absorption in the crystal on the line width can be ignored and the main contribution to the line broadening of the backward PXR in extremely Bragg geometry is made by the multiple scattering process of relativistic electrons in crystal. 
\section{Numerical results and discussion}\label{sec4}
\begin{figure}
  \centering
  \includegraphics[width=\linewidth]{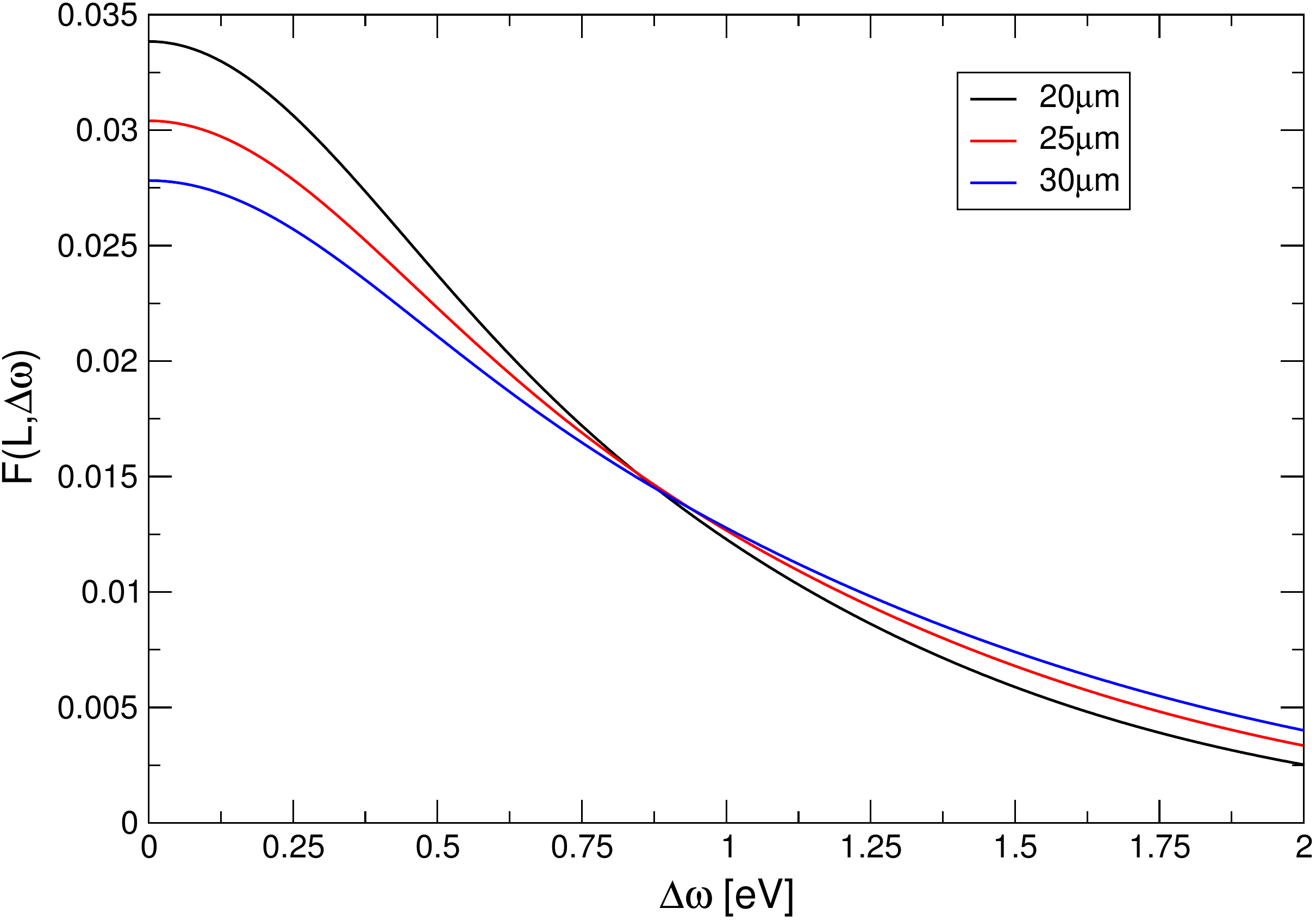}
  \caption{Numerical results of the multiple scattering effect on the  line width $\Delta\omega$ of the backward PXR according to (\ref{eq5}) produced by relativistic electrons of the energy of $E=30$ MeV traversing a Si crystal of the varying thicknesses $20\mu m\leq L\leq 30\mu m$ (see Table \ref{tab1}).}
  \label{fig3}
\end{figure}

Formula (\ref{eq2}) is appropriate for a parametric consideration of the multiple
scattering effect of relativistic electrons on the line width of the backward PXR.
But it is less suitable for numerical analysis from the wall time point of view.
Therefore, to investigate numerically the influence of the multiple scattering on the line width, one has to modify formula (\ref{eq2}) as in the following.\\
Taking $p\equiv2L\Delta\omega$ and $s\equiv\alpha^{2}\sigma^{2}$ as new numerical dimensionless quantities and using Euler representation for cosine function, one gets the following formula for $F(L,\Delta\omega)$
\begin{equation}
\begin{split}
\rm
F(L,\Delta\omega)= 2\int_{0}^{1}\frac{\sqrt{\pi}}{2\sqrt{sz}}\,e^{-\frac{p^{2}}{4sz}}\Bigg[ erf\Bigg(\frac{ip+2sz^{2}}{2\sqrt{sz}}\Bigg)\\
\rm
-erf\Bigg(\frac{ip-2sz^{2}}{2\sqrt{sz}}\Bigg)\Bigg]dz,
\end{split}
\label{eq5}
\end{equation}
where the line width $\Delta\omega$ is included in the arguments of complex Error and  exponential functions (In Eq.(\ref{eq5}), $\rm i$ is the imaginary unit).\\
The Eq. (\ref{eq5}) depends on the parameters such as thickness $L$ of crystal, energy $E$ of relativistic electrons, incident angle $\psi$ of electrons to the crystallographic axis ($\psi$ must be more than $\psi_{c}$) and  $\omega_{n}$. 
For concrete practical application of formula (\ref{eq5}) under the conditions (\ref{eq4}), we consider the line width of the backward PXR produced by relativistic electrons of different $E=30$ MeV and $E=50$ MeV energies in a single Si crystal of varying thicknesses. The radiation length for Si is $L_{R}=9.37$ cm\cite{PDG} and we use $\omega_{n}\approx\frac{4\pi}{\sqrt{3}d_{Si}}n$ with $n=3$, where $d_{Si}=5.43\times10^{-8}$cm \cite{KIT}  (for absorption length of radiated photons in the energy range of $7-8$ KeV in Si, see X-Ray Attenuation Length \cite{BERK}).\\
For above $\omega_{n}$, the thickness of Si crystal and the values of the energy of relativistic electrons
are chosen so that the condition $\alpha\gg 1$ and $\sigma_{x}\gg 1$ of application of Eq. (\ref{eq1}) together with (\ref{eq4}) could be fulfilled. Then the optimal value for the energy of relativistic electrons is started from $30$ MeV to retain particles in relativistic domain. The crystal thickness (first column of Table \ref{tab1}) is in the range of $20\mu m\leq L\leq 30\mu m$ ($L<L_{a}$) in order to fulfil both conditions $\alpha\gg 1$ and $\sigma_{x}\gg 1$ (second and third columns of Table \ref{tab1}).\\
\begin{table}[ht]
\caption{Approximated and exact values of the line width of the backward PXR (fourth and fifth columns) for electrons of the energy of $E=30$ MeV traversing a Si crystals of varying thicknesses (first column). The condition $\alpha\sigma_{x}\gg 1$ is fulfilled.}
\centering
\begin{tabular}{c c c c c}
\hline\hline
L[$\mu m$] & $\alpha$ & $\sigma_{x}$ & $\Delta\omega_{app}$[eV](\ref{eq3}) & $\Delta\omega_{ex}$[eV]\\ [0.3ex]
\hline
20 & 22.24 & 8.14 & 1.81 & 1.52 \\
25 & 19.64 & 13.05 & 2.05 & 1.68 \\
30 & 17.75 & 19.17 & 2.26 & 1.83  \\ [0.3ex]
\hline
\label{tab1}
\end{tabular}
\end{table}
\begin{table}[ht]
\caption{Same as in Table \ref{tab1} but for electrons of the energy of $E=50$ MeV traversing Si crystal of different thicknesses (first column).}
\centering
\begin{tabular}{c c c c c c}
\hline\hline
L [$\mu m$] & $\alpha$ & $\sigma_{x}$ & $\Delta\omega_{app}$[eV](\ref{eq3}) & $\Delta\omega_{ex}$[eV]\\ [0.5ex]
\hline
35 & 27.17 & 9.55 & 1.48 & 1.21 \\
40 & 25.23 & 12.66 & 1.59 & 1.29 \\
45 & 23.64 & 16.22 & 1.70 & 1.36  \\
50 & 22.30 & 20.26 & 1.80 & 1.44  \\ [1ex]
\hline
\label{tab2}
\end{tabular}
\end{table}	
The optimal upper limit for energy of electrons is $50$ MeV, since at the chosen thicknesses (first column of Table \ref{tab2}), the conditions (\ref{eq4}) (second and third columns of Table \ref{tab2}) are still fulfilled for photons in the $7-8$ KeV energy range.\\ For $100$ MeV electrons traversing Si crystal of the thickness $10\mu m\leq L \leq 60\mu m$, the values of $\sigma_{x}$ are in the range of $0.17\leq\sigma_{x}\leq 7.4$. Therefore, the condition $\sigma_{x}\gg 1$ can not be fulfilled. 
It seems that more thicker crystal $L>60\mu m$ could be used to increase $\sigma_{x}$. But such range of crystal thickness is not suitable, since the absorption of radiated photons parasitically influences on the line width of the backward PXR and  the first part of conditions (\ref{eq4}) can not be fulfilled. Thus, the upper limit of energy for electrons is set to $50$ MeV.\\
Based on the used $\omega_{n}$ together with the selected ranges for energy of electrons and the corresponding values of incident angle $\psi$ of electrons and appropriate thickness of Si, the numerical calculations of the influence of multiple scattering of $E=30$ MeV and $E=50$ MeV relativistic electrons on the line width of  the backward PXR in extremely Bragg geometry are shown in Figs. \ref{fig3} and \ref{fig4}.\\
Let us discuss the obtained curves in both figures by the example of $L=20\rm\mu m$ for $E=30$ MeV electrons  (Fig. \ref{fig3}).
The maximum value of $F(L,\Delta\omega)$ for $L=20\rm\mu m$ at chosen energy of electron in Fig. \ref{fig3} is $\simeq 3.3\times 10^{-2}$. Dividing it by two and finding the corresponding value of $\Delta\omega$ for it on the abscissa of Fig. \ref{fig3} and multiplying the obtained $\Delta\omega$ by two, we get $\Delta\omega_{ex}=1.52\rm\,eV$ for the line width of the backward PXR produced by $E=30$ MeV relativistic electrons traversing Si crystal of thickness $L=20\rm\mu m$ taking into account the scattering process in the absence of photo absorption, where $\Delta\omega_{ex}$ means the exact calculated line width by Eq.\ref{eq5} . This value is shown in the fifth column of Table \ref{tab1}.  
The same value of $\Delta\omega_{app}$ for $L=20\rm\mu m$, but according to the formula (\ref{eq3}) is shown in the fourth column of Table \ref{tab1}.\\
There is a small discrepancy $\Delta$ between the values of the line width for $L=20\rm\mu m$ in the fourth and fifth columns of Table \ref{tab1}, where $\Delta=|\Delta\omega_{app}-\Delta\omega_{ex}|$. This is because the line width in the fourth column $\Delta\omega_{app}$ is obtained by the approximate formula (\ref{eq3}), whereas $\Delta\omega_{ex}$ in fifth column is extracted from  the exact calculation of (\ref{eq5}) in Fig. \ref{fig3}.
\\
The above discussion concerning the numerical result of the line width at $L=20\rm\mu m$ in Fig. \ref{fig3} can be given for other thicknesses in Fig. \ref{fig3} and Table \ref{tab1}, as for corresponding values in Fig. \ref{fig4} and Table \ref{tab2}.\\
This is very important to note that the natural line width of the backward PXR produced by $E=30$ MeV relativistic electrons at $L=20\rm\mu m$  in extremely Bragg geometry under the condition of the absence of photo absorption is about $\Delta\omega\sim 10$ meV to within the order of magnitude. 
This value is much smaller than the line width of the backward PXR of the same thickness, but affected by the multiple scattering effect $\Delta\omega_{ex}=1.52$ (fifth column in Table \ref{tab1}). This difference is of significant for all thicknesses at both $E=30$ MeV and $E=50$ MeV energies. It means that the multiple scattering of relativistic electrons traversing a single crystal  makes a significant contribution to the line width of the backward PXR in the absence of photo absorption.
\section{Conclusions}
The influence of the multiple scattering of low energy $30$ and $50$ MeV relativistic electrons on the line width $\Delta\omega$ of the backward PXR in extremely Bragg geometry in a Si crystal at varying thicknesses has been considered. The conditions, when the photo absorption effect on the line width is negligible, has been obtained.\\
The obtained numerical values of $\Delta\omega$ show that the scattering process makes the important contribution to the line width of the backward PXR in the absence of photo absorption.\\
The line width of the backward PXR in extremely Bragg geometry at low energy relativistic electrons in experiment \cite{FREUD1} was not measured, although a boost of intensity by a factor of two was observed. As far as we know,  apart from the measurement of the line width of the backward PXR at  high energies of electrons\cite{BACKE1}, no any experiment has been performed yet on the measurement of the line width of the backward PXR in extremely Bragg geometry, produced by low energy relativistic  electrons in thin crystals. The experiment could make the influence of the multiple scattering effect on the line width of the backward PXR more clear, since this is a still open question.
\section{Acknowledgements}
The author is grateful to N.F. Shul'ga from the Kharkov Institute of Physics and Technology (Ukraine) for discussion concerning the work and reading the manuscript.\\
\begin{figure}
  \centering
  \includegraphics[width=\linewidth]{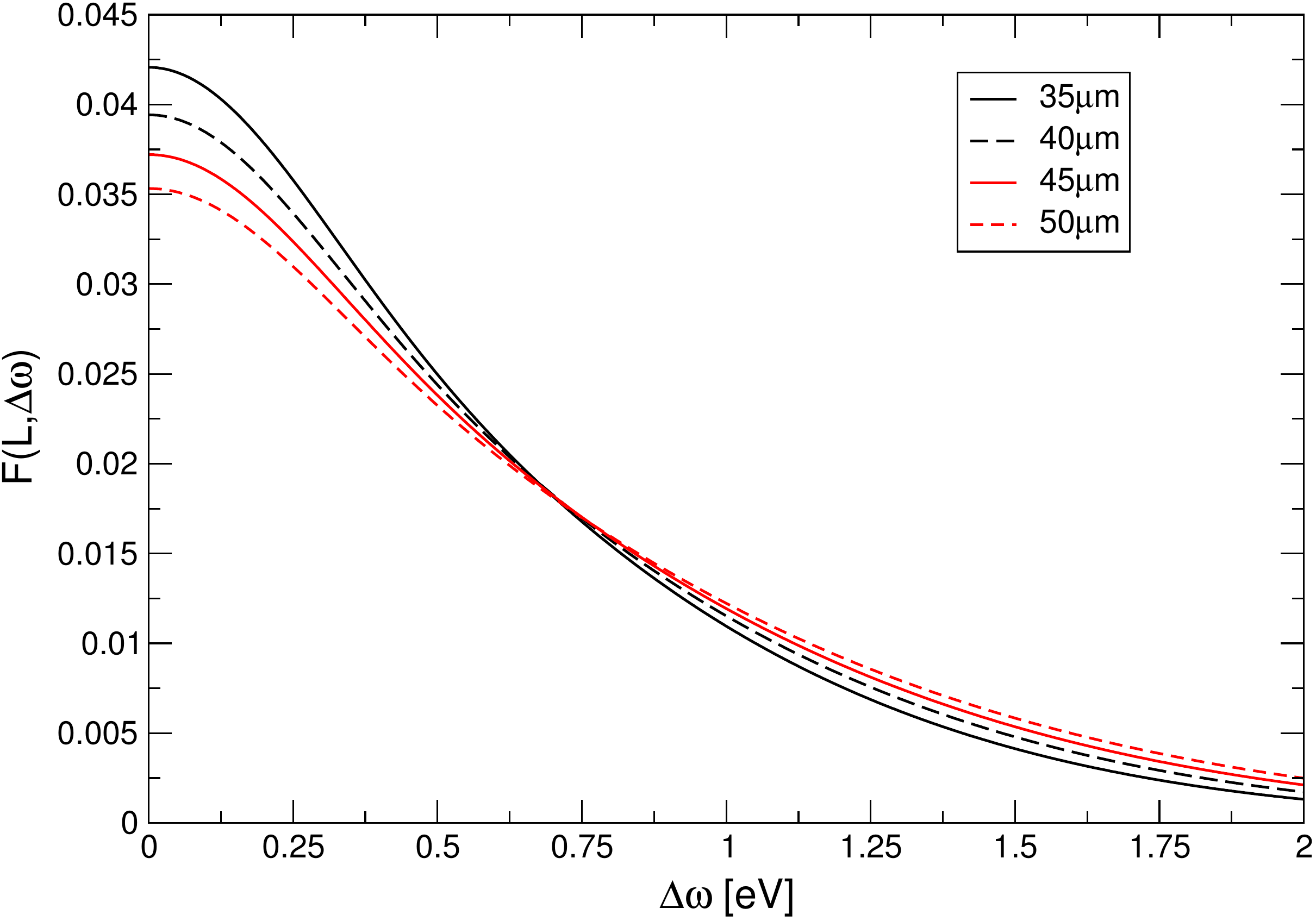}
  \caption{Same as in Fig. \ref{fig3} but for electrons of the energy of $E=50$ MeV traversing a Si crystal of the varying thicknesses $35\mu m\leq L\leq 50\mu m$ (see Table \ref{tab2}).}
  \label{fig4}
\end{figure}
The author would like in particular to express
his gratitude to Tanaji Sen from the Fermi National Lab (USA) for fruitful discussions concerning this work
and useful recommendations during reading the manuscript.\\
Figs. \ref{fig1} and \ref{fig2} have been prepared in the free and open-source software "LibreOffice Draw"\cite{LOD}.\\
The numerical results in Figs. \ref{fig3} and \ref{fig4} have been calculated in the free and open-source software  Sage\cite{SAGE} at the Parallel Computing Laboratory (2216) of the Department of Physics at the Razi University of Kermanshah using GSL (GNU Scientific Library) and NumPy packages..\\
Active administrative support from Dr. Ardeshir Rabeie together with the
financial support from the Department of Physics for installing and establishment of the Parallel Computing Laboratory, are gratefully acknowledged.
\section*{References}
\bibliography{mybibfile}
\end{document}